\documentclass[11pt]{spie} 
\usepackage[T1,T2A]{fontenc}
\usepackage[english,russian]{babel}
\usepackage{setspace}

\usepackage[utf8]{inputenc} 

\usepackage{geometry} 
\usepackage{graphicx} 
\usepackage{epsfig}
\usepackage{miniplot}

\usepackage{booktabs} 
\usepackage{array} 
\usepackage{paralist} 
\usepackage{verbatim} 
\usepackage{amsmath}
\usepackage{subfig} 

\usepackage{fancyhdr} 
\usepackage{sectsty}
\allsectionsfont{\sffamily\mdseries\upshape} 

\usepackage[nottoc,notlof,notlot]{tocbibind} 
\usepackage[titles,subfigure]{tocloft} 

\numberwithin{equation}{section}

\title{Classical Equation of Electromagnetic Field in the Higgs Boson Field and Estimation on the Static Electrical Polarizability of Leptons}
\author{Il-Gwang Kim\supit{a}, Tae-Song Kim\supit{b}, Chol-Man Ri\supit{c} and Song-Jin Im\supit{d} \skiplinehalf
\supit{a} Natural Science Institute, Kim Il Sung University, Pyongyang, DPRK\skiplinehalf
\supit{b} Faculty of Energy Science, Kim Il Sung University, Pyongyang, DPRK\skiplinehalf
\supit{c} E-Library, Kim Il Sung University, Pyongyang, DPRK\skiplinehalf
\supit{d} Faculty of Physics, Kim Il Sung University, Pyongyang, DPRK}
\authorinfo{Kim Tae-Song: E-mail: ryongnam13@yahoo.com}

\date{} 

\begin{document}
\maketitle
\selectlanguage{english}

\begin{abstract}
In our paper we derived the classical motion equation of electromagnetic field in space with Higgs field and by means of it discussed the distributions of charge and current formed when the static electrical and magnetic fields are interacting with the spherically symmetrical Higgs field, and predicted the electrical polarizability of electron.
\end{abstract}


\section{Classical motion equation of electromagnetic field under Higgs field}

\begin{figure}[!ht]
\begin{center}
\includegraphics[clip=true,scale=1]{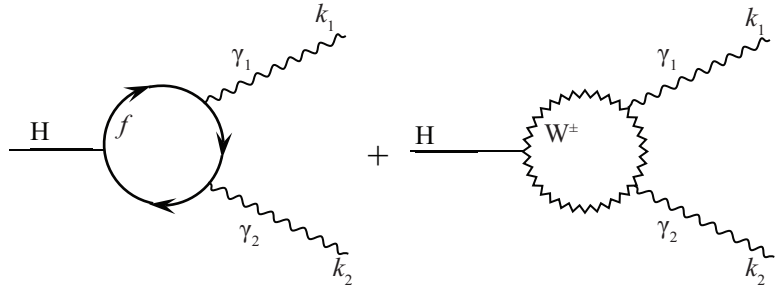}
\caption{\label{fig:one-vol}Feynman Diagram on interaction between Higgs field $H$ and electromagnetic field $\gamma$ through the rings of the main Fermions $f$ and the immediate vector bosons $W^+$, $W^-$}
\end{center}
\end{figure}

As already studied by previous workers, Feynman diagram(Fig. \ref{fig:one-vol}) which represents the interaction between electromagnetic field and Higgs field and the corresponding Lagrangian $L_{H_{\gamma\gamma}}$(Eq. \ref{eq:1}) is given as \cite{ref1, ref2, ref3}

\begin{equation}
\label{eq:1}
\mathcal{L}_{H_{\gamma\gamma}}=\frac{\alpha F}{8\pi\eta}H(x)\left(F_{\mu\nu}(x)\right)^2
\end{equation}

Here $\gamma$ is the fine-structure constant, $\eta=(G\sqrt{2})^{-1/2}=246GeV$, and considering the virtual ring particle $F$ is calculated as
\begin{equation}
\label{eq:2}
F=\sum_n F_n Q_n^2 - F_I
\end{equation}
, where $n$ means the charged virtual ring particles – leptons ($e$, $\mu$, $\tau$), quarks and bosons $W^+$, $W^-$, $Q_n$ represent charges of these particles , and $F_n$ are calculated through the equations (1-3), (1-4).
\begin{equation}
\label{eq:3}
\left\lbrace
\begin{split}
    F_{\frac{1}{2}}&=-2\beta[(1-\beta)\chi^2+1]; \text{ $\frac{1}{2}$ spin particle}\\
    F_1&=2+3\beta+3\beta(2-\beta) \chi^2; \text{1 spin particle}\\
    F_I&; \text{contribution of strong interaction}
\end{split}
\right.
\end{equation}
Here
\begin{equation}
\label{eq:4}
\left\lbrace
\begin{split}
    \beta&=\frac{4m^2}{m^2_H}\\
    \chi&=\arctan{\frac{1}{\sqrt{\beta-1}}}; \beta>1\\
    \chi&=\frac{1}{2}\left[i \ln{\frac{1+\sqrt{1-\beta}}{1-\sqrt{1-\beta}}}+\pi\right]; \beta <=1
\end{split}
\right.
\end{equation}
$m$ is the mass of the virtual ring particle, $m_H$ is the mass of Higgs particle and Lagrangian which describe the motion of electromagnetic field in space with Higgs field is given as
\begin{equation}
\label{eq:5}
L=L_0+L_{H_{\gamma\gamma}}
\end{equation}
, where $L_0$ is Lagrangian of free electromagnetic field, i.e. 
\begin{equation}
\label{eq:6}
L_0=-\frac{1}{4}F_{\mu\nu} F^{\mu\nu}
\end{equation}
In eq. (\ref{eq:1}) introducing the representation
\begin{equation}
\label{eq:7}
U(x)=-\frac{\alpha F}{8\pi\eta}H(x),
\end{equation}
eq. (\ref{eq:5}) can be written as
\begin{equation}
\label{eq:8}
L=-\frac{1}{4} \left(1+4U(x)\right) F_{\mu\nu}F^{\mu\nu}.
\end{equation}
Also taking account for
\begin{equation}
\label{eq:9}
\frac{\partial(F_{\mu\nu}F^{\mu\nu})}{\partial(\partial^\mu A^\nu)} =2F_{\mu\nu},
\end{equation}
the classical motion equation of electromagnetic field is given as
\begin{equation}
\label{eq:10}
\frac{\partial}{\partial x_\mu} \left[\left(1+4U(x)\right) F_{\mu\nu} \right]=0.
\end{equation}
This equation can be rewritten as 
\begin{equation}
\label{eq:11}
\frac{\partial F_{\mu\nu}}{\partial x_\mu}=-\frac{4}{1+U(x)}\frac{\partial U(x)}{\partial x_\mu}F_{\mu\nu}.
\end{equation}
Using the representations of electrical field vector and magnetic field vector, we can obtain the followings equation system.
\begin{equation}
\label{eq:12}
\left\lbrace
\begin{split}
div \mathbf{E}&=-\left( grad \left( \ln{\left(1+U(x) \right)} \right) \cdot \mathbf{E} \right)  \\
\frac{\partial \mathbf{E}}{\partial t}- rot \mathbf{H}&=grad\left(\ln{\left(1+4U(x)\right)}\right )\times\mathbf{H}\\
rot\mathbf{E}&=-\frac{\partial \mathbf{H}}{\partial t}\\
div\mathbf{H}&=0
\end{split}
\right.
\end{equation}
As well-known, if sources of electromagnetic field – charge and current exist, eq. (\ref{eq:2}) is changed as
\begin{equation}
\label{eq:13}
\left\lbrace
\begin{split}
div \mathbf{E}&=-\rho \\
\frac{\partial \mathbf{E}}{\partial t}- rot \mathbf{H}&=-\mathbf{j}\\
rot\mathbf{E}&=-\frac{\partial \mathbf{H}}{\partial t}\\
div\mathbf{H}&=0
\end{split}
\right.
\end{equation}
, where $\rho$ and $j$ are the charge density, the current density, respectively.
Comparing eq. (\ref{eq:13}) with eq. (\ref{eq:12}), we can find that the following new type of charge and current sources occur by interaction between Higgs and electromagnetic fields.
\begin{equation}
\label{eq:14}
\rho=-\left[ grad           
 \left(
 \ln{\left(1+U(x) \right)}
 \right)
\cdot\mathbf{E}
\right]
\end{equation}
\begin{equation}
\label{eq:15}
\mathbf{j}=-\left[
grad\left(
\ln{\left(1+4U(x)\right)} 
\right)\times\mathbf{H}
\right]
\end{equation}

\begin{figure}[!ht]
\begin{center}
\includegraphics[clip=true,scale=1]{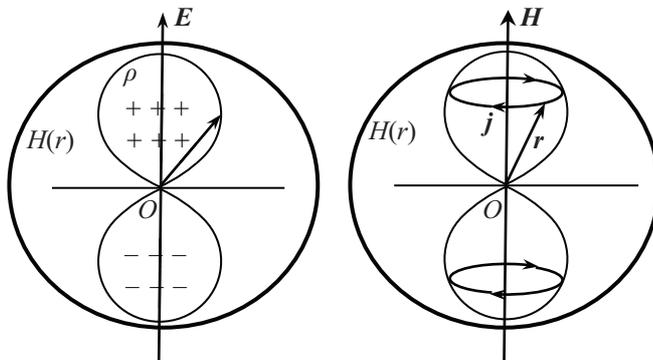}
\caption{\label{fig:two-vol}Spherically symmetric Higgs field and distributions of charge and current density induced by uniform electrical field $\mathbf{E}$ and uniform magnetic field $\mathbf{H}$ when $\frac{d}{d r}\left(\ln{\left(1+U(r)\right)}\right)<0$.}
\end{center}
\end{figure}

In order to obtain the clear image of space distribution of space-charge density $\rho$ and space-current density $\mathbf{j}$, first as the simplest case we assume that spherical symmetrical Higgs field $H(r)$ distributed round a given center, static uniform electrical field $\mathbf{E}$ and uniform magnetic field $\mathbf{H}$ exist.
Then eqs. (\ref{eq:14}) and (\ref{eq:15}) can be written as
\begin{equation}
\label{eq:16}
\rho=-\frac{1}{r}\frac{d}{d r}\left(\ln{\left(1+U(r)\right)}\right)(\mathbf{r}\cdot\mathbf{E})
\end{equation}
\begin{equation}
\label{eq:17}
\mathbf{j}=-\frac{1}{r}\frac{d}{d r}\left(\ln{\left(1+U(r)\right)}\right)\left[\mathbf{r}\times\mathbf{H}\right]
\end{equation}
Distributions of Charge density and current density which are taken from these equations can be seen from Fig \ref{fig:two-vol}.

\begin{figure}[!ht]
\begin{center}
\includegraphics[clip=true,scale=1]{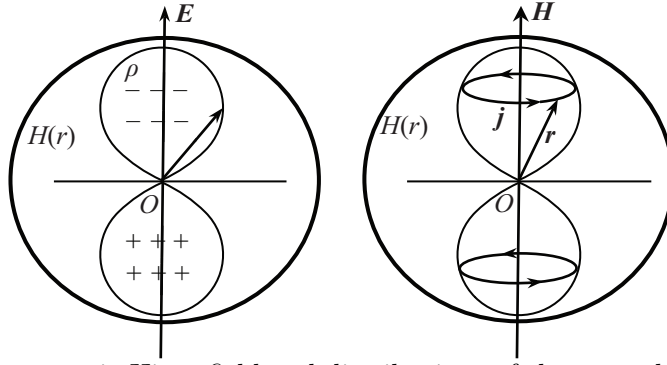}
\caption{\label{fig:three-vol}Spherically symmetric Higgs field and distributions of charge and current density induced by uniform electrical field $\mathbf{E}$ and uniform magnetic field $\mathbf{H}$ when $\frac{d}{d r}\left(\ln{\left(1+U(r)\right)}\right)>0$.}
\end{center}
\end{figure}

As you see from Fig \ref{fig:two-vol}, when electrical field and magnetic field exist, charge and current distributions of dumbbell shape, of which axes are electrical field vector and magnetic field vector passing across the center of spherically symmetric Higgs field, are induced and occur in direction which makes the external electrical and magnetic fields weaker when Higgs field decreases with the increment of distance, i.e. when $\frac{d}{d r}\left(\ln{\left(1+U(r)\right)}\right)<0$. In other words Higgs field acts as “dielectric body” or “diamagnetic” in this case.
On the contrary when strength of Higgs field increases with r, opposite phenomenon to one above mentioned presents (Fig \ref{fig:three-vol}).

Since generally the strength of Higgs field built around spinor is proportional to its mass and mass of Higgs field particle is very large as 125GeV, it decreases rapidly as the distance from the particle increases.

In other words spinor particle appears to put on an ultrathin cloth of Higgs field.
Since such Higgs field exists, particle will take additional electrical polarizability and magnetic polarizability if it enters electric field and magnetic field.

\section{Electrical Polarizability of Main Particles of Standard Gauge Model}

The wave function of the main particles $e,\mu,\tau$ of Standard Gauge Model, quarks and fermions is expressed as $\Psi(x)$. Then Lagrangian of interaction between these particles and Higgs field can be written as 
\begin{equation}
\label{eq:201}
\mathcal{L}_{H_{\bar{\Psi}\Psi}}=\frac{m_\Psi}{\eta}\bar{\Psi}(x)\Psi(x)H(x)
\end{equation}
In order to obtain the space-time distribution of Higgs field built around spinor particle properly, we have to solve the complicated nonlinear partial differential equations with the nonlinear terms as $H^2 (x)$ and $H^3 (x)$, but for the sake of brief discussion we will use the approximate solution of Yukawa field shape as follows:
\begin{equation}
\label{eq:202}
H(r)=-\frac{m_\Psi}{4\pi\eta} \frac{e^{-m_H r}}{r}
\end{equation}
Equation (\ref{eq:202}) reflects rather properly the strength of Higgs field in some long distance from Higgs field source. As you see from this equation, Higgs field almost vanishes over the distance of $\frac{1}{m_H}$  from the source and is concentrated within the distance shorter than that.
Now for the sake of convenience we will introduce the representation 
\begin{equation}
\label{eq:203}
f(r)=\ln{\left(1+4U(r)\right)}.
\end{equation}
Calculating by using Eq. (\ref{eq:16}) about the charge density which appears when external electrical field $E$, electrical dipole is derived as
\begin{equation}
\label{eq:204}
d=-\int \frac{df(r)}{d r} (\mathbf{r}\cdot\mathbf{E}) \frac{\mathbf{r}}{r} dV
\end{equation}
, or in spherical coordinates
\begin{equation}
\label{eq:205}
\mathbf{d}=-\int_0^\infty\int_0^\pi\int_0^{2\pi} \frac{df(r)}{dr} (\mathbf{r}\cdot\mathbf{E}) \frac{\mathbf{r}}{r} r^2  \sin{\theta} dr d\theta d\varphi.
\end{equation}

This is represented in spherical coordinates where the origin point is the position where Higgs field source – a particle exists, and then the sum of $x$ and $y$ components of electrical dipole vanishes and only $z$ component remains, i.e.:
\begin{equation}
\label{eq:206}
d=\left(-\frac{4\pi}{3} \int_0^\infty dr\cdot r^3 \frac{df(r)}{dr}\right)E
\end{equation}
The quantity within parentheses is the electrical polarizability $\beta$ by Higgs field built around fermion. In order to estimate the magnitude of the electrical polarizability $\beta$, we will first replace $r$ with the dimensionless quantity $\xi=m_H r$ and then substitute eq. (\ref{eq:202}) about Higgs field for eq. (\ref{eq:203}) about $f(r)$.
Then we can rewrite eq. (\ref{eq:7}) about $U(r)$ as
\begin{equation}
\label{eq:207}
U(\xi)=C_0  \frac{e^{-\xi}}{\xi}
\end{equation}
, where $C_0$ is
\begin{equation}
\label{eq:208}
C_0=\frac{\alpha Fm_H m_\Psi}{32\pi^2 \eta^2}
\end{equation}
, and 
\begin{equation}
\label{eq:209}
\frac{df(\xi)}{d\xi}=-4C_0\frac{\xi+1}{\xi(\xi e^{\xi}+4C_0)}.
\end{equation}
Calculating the equation of the polarizability $\beta$ by using eq. (\ref{eq:209}), we find 
\begin{equation}
\label{eq:210}
\beta=-\frac{4\pi}{3}\int_0^\infty dr r^3  \frac{df(r)}{dr}=C_1 \int_0^\infty d\xi \frac{\xi^2+\xi^3}{\xi(e^\xi+4C_0 ) }
\end{equation}
, where 
\begin{equation*}
C_1=\frac{\alpha Fm_\Psi}{6\pi m_H^2 \eta^2 }
\end{equation*}
On the other hand, as $\frac{m_H}{\eta}\approx 0.5 $, $\frac{m_\Psi}{\eta}\approx 2\cdot 10^{-6}$, $\alpha=\frac{1}{137}$, $F\approx tens$, in case of the electron $C_0 \approx 10^{-10}$ is very small.
Evaluating the integral (\ref{eq:210}) from this, it is about 3 and finally 
\begin{equation}
\label{eq:211}
\beta \approx \frac{\alpha Fm_\Psi}{2\pi m_H^2 \eta^2 }.
\end{equation}
Calculating the polarizability $\beta$ due to Higgs field built around the fermion, we can find that we have to be accurately aware of mass $m_H$ of Higgs field particle as well as mass $m_\Psi$ of a given spinor particle. Recently the fact that mass of Higgs boson $m_H\approx 125GeV$ is confirmed both theoretically and experimentally \cite{ref4, ref5, ref6}.
Along with this, substituting masses of $e,\mu$ and $\tau-$mesons; $m_e=5.11\times 10^{-4} GeV$, $m_\mu=0.1057GeV$, $m_\tau=1.784GeV$ and the value of $F$, computed numerically by using equations (\ref{eq:2}) $\sim$ (\ref{eq:4}) for equation (\ref{eq:211}), the polarizabilities $\beta$ of $e,\mu$ and $\tau$ are calculated as
\begin{equation}
\label{eq:212}
\begin{split}
\beta^{(e)} \approx 4.5\cdot 10^{-57} cm^3\\
\beta^{(\mu)}\approx 1.0\cdot 10^{-54} cm^3\\
\beta^{(\tau)} \approx 1.6\cdot 10^{-53} cm^3
\end{split}
\end{equation}
As already known, the polarizabilities of atoms are similar to their volumes approximately. On the contrary, if the polarizability of an atom is known, its diameter can be estimated approximately.
With this respect, for instance, electron magnitude is estimated about $10^{-19} cm$.
On the other hand, as static polarizability must be real, it is required that variable F in equation (\ref{eq:211} must be real and then $\mathrm{Im} F=0$.
From this we can search numerically the mass $m_H$ of Higgs particle satisfying $\mathrm{Im} F=0$. 
We will stress that the mass $m_H$ of Higgs boson is between $120 \sim 164GeV$ within a certain error by numerically computing.

\section*{Conclusions}
In our paper we derived the classical motion equations of electromagnetic field under Higgs field using the simplest ring diagram of interaction between Higgs field and electromagnetic field and as a simple application of it we revealed the charge polarization and magnetic polarization induced when spherically symmetric Higgs field exist in the static electrical and magnetic fields.
Even though the main fermions of Standard Gauge Model don’t consist of other components, we also derived the evaluation of the electrical polarizability due to Higgs field shell built around them and from it calculated leptons’ polarizabilities approximately.

\acknowledgements
I thank Kim Nam-Hyok for careful discussion on this paper.

\selectlanguage{english}

\end{document}